\newcommand{\be}{\begin{equation}}
\newcommand{\ee}{\end{equation}}
\newcommand{\bea}{\begin{eqnarray}}
\newcommand{\eea}{\end{eqnarray}}
\newcommand{\ba}[1]{\begin{array}{#1}}
\newcommand{\ea}{\end{array}}

\documentclass[twocolumn,secnumarabic,amssymb,nobibnotes,nofootinbib,aps,prl,showpacs]{revtex4}
\usepackage{epsfig}
\usepackage{amssymb}
\begin{document}
\setlength{\topmargin}{-0.05in}

\title{Manipulating Higher Partial-Wave Atom-Atom Interaction by Strong Photoassociative Coupling }
\author{B. Deb$^{1,2}$ and J. Hazra$^1$}
\affiliation{$^1$Department of Materials Science, and
$^2$Raman Center for Atomic, Molecular and Optical Sciences,
Indian Association
for the Cultivation of Science,
Jadavpur, Kolkata 700032, India.}


\def\zbf#1{{\bf {#1}}}
\def\bfm#1{\mbox{\boldmath $#1$}}
\def\hf{\frac{1}{2}}

\begin{abstract}
We show that it is possible to change not only s-wave  but also higher partial wave atom-atom interactions in cold collision in the presence of relatively intense laser fields tuned near a photoassociative transition.

\end{abstract}

 \maketitle

Ability to control particle-particle interaction is important for exploring quantum physics of many-particle systems in various interaction regimes.
 Ultracold atoms offer a unique opportunity for such explorations with unprecedented control over atom-atom interaction.
  There are two methods of manipulating interaction in cold atoms.
  The most popular one is magnetic field Feshbach resonance (MFR)
  \cite{tiesingaPRA47}
   which has been extensively used to tune s-wave scattering length
   over a wide range. This has facilitated the recent
   demonstration
   of s-wave fermionic superfluidity in strongly interacting
   atomic gases \cite{Ketterle}. In fact, MFR  has become an
   essential tool in experimental investigations on
the effects of large s-wave scattering length on the properties of
atomic Fermi gases \cite{fermi} and Bose-Einstein condensates (BEC)
\cite{bose}. The other method of modifying atomic interaction is
optical Feshbach resonance (OFR) proposed by Fedichev {\it et al.}
\cite{fedichevPRL771} and implemented in recent experiments
\cite{fatemiPRL2002,theis,EnomotoPRL101}. While MFR relies on
magnetic effects of Zeeman and hyperfine interactions, OFR uses
off-resonant continuum-bound optical dipole transitions. In the case
of resonance or near-resonance, OFR can lead to photoassociation
(PA) \cite{parmp} of two atoms into an excited molecule. Recently,
p-wave MFR \cite{ZhangPRA70} in fermionic atoms  has  been observed.
Enhanced scattering in higher partial waves by magnetic-field
induced dissociation of Feshabch molecule has been shown
\cite{rempePRA72}. There is a  proposal \cite{you} for generating
anisotropic interaction by static electric field. Both the methods
of magnetic and optical Feshbach resonances  are so far primarily
used to tune s-wave scattering length in ultracold atoms. To go
beyond s-wave physics of cold atoms, it is now essential to devise
methods of controlling p-, d- and other higher partial-wave
interactions. This is particularly important for testing  models of
unconventional superconductivity or superfluidity in atomic Fermi
gases. Superfluidity and superconductivity are related phenomena.
Conventional low temperature superconductivity can be explained by
Bardeen-Cooper-Schrieffer theory which is based on s-wave
Cooper-pairing. It is assumed that higher partial-wave interactions
can lead to unconventional and high temperature superconductivity.
Studies on Fermi superfluidity in cold atomic gases with
controllable p- and d-wave interactions will help us to develop new
insight about high temperature superconductivity which requires a
proper theoretical understanding.

Here we show that it is possible to change not only $\ell = 0$
(s-wave)  but also nonzero partial-wave scattering amplitudes of two
cold atoms by OFR with a relatively intense laser field.  At low
energy, the light-shift (or Stark-shift) due to laser-induced
free-bound coupling can greatly exceed the spontaneous as well as
stimulated line widths  of excited molecular state. An intense PA
laser can set in two photon processes in which one photon will cause
PA transition from continuum to bound level and another photon will
induce stimulated transition back to the continuum. If the
light-shift largely  exceeds the stimulated line width, then even
when PA laser is tuned near the unperturbed (without Stark shift) excited molecular level, the
formation of excited molecule becomes unlikely due to large
light-shift. In such a situation, s-wave scattering wave function
can be made to couple to p-wave or even d-wave scattering wave
functions depending on the coupling of the molecular axis with the
electronic orbital and spin angular momentum.  Furthermore,  it is
possible to enhance $\ell \ne 0$
 partial wave scattering amplitudes with multiple strong laser fields
 causing continuum-bound PA coupling with appropriate rotational states of an excited vibrational level
 as illustrated in Fig.1.
 \begin{figure}
 \includegraphics[width=4.25in]{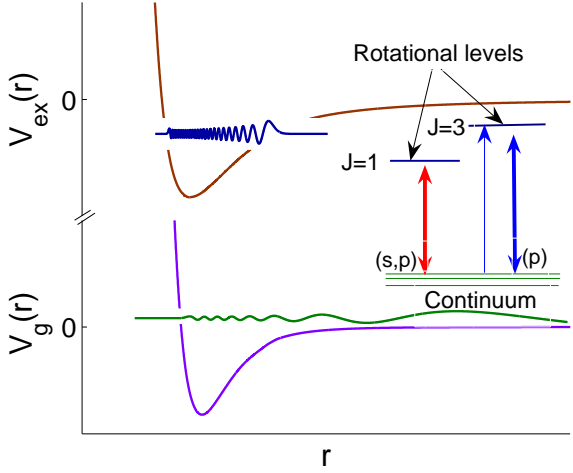}
 \caption{A schematic diagram of ground and excited potentials, scattering and bound states,
 rotational levels and relevant PA transitions for modifying $\ell \ge 0$ partial wave scattering amplitudes.
 An intense laser (double arrow red line on the left) is tuned near $J = 1$ rotational
 state
 (of a particular vibrational level $v$) which can be accessed by PA transitions from s-wave
 and also at least the next nonzero (p-wave) scattering state.
 In the strong coupling dispersive regime with large light shift (see text),
 p-wave scattering amplitude will get modified due to its indirect coupling with s-wave scattering state.
 This modification can be probed
 by sending a weak probe laser (single arrow blue line) resonant with $J = 3$ PA transition.
 The modification of p-wave can further be enhanced
 by applying another intense laser (double arrow blue line on the right) tuned near $J = 3$ PA transition.}
 \end{figure}
In this context,  recent experimental results \cite{EnomotoPRL101}
on the intense
 laser field PA of ytterbium  may be of
 relevance. In molecular bound-bound spectroscopy, it is
 known that  the
 rotational states of a molecule can be excited by intense laser fields
 \cite{seidmanJCP103}, but exciting higher partial waves in
 continuum states by continuum-bound PA spectroscopy has not been
 considered so far.

Let us consider that the scattering state of  collision energy $E
= \hbar^2 k^2 /(2\mu)$ (where $\mu$ is the reduced mass) of two
colliding ground state atoms is coupled to an excited
 molecular state characterized by $v$ vibrational
and $J$ rotational quantum numbers.  The electronic orbital ($L$)
and spin ($S$) angular momentum  of the excited diatom are coupled
to the diatomic axis according to either Hund's case (a) or (c).
In Hund's case (a),  $\Lambda$ and $\Sigma$ which are the
projections of ${\mathbf L}$ and ${\mathbf S}$, respectively, on the internuclear axis
are two good quantum numbers and so  is their sum $\Omega =
\Lambda
 + \Sigma$. In Hund's case (c), the projection $\Omega$ of
 the total electronic angular momentum $\mathbf{J}_e = \mathbf{L} + \mathbf{S}$ is a good quantum number.
   The angular state of diatom can be written as
  $ \mid J \Omega M \rangle = i^J \sqrt{\frac{2 J + 1}{8\pi^2}} {\cal D}^{(J)}_{M \Omega }(\hat{r})
  $
   where $M$ is the z-component of J in the  space-fixed coordinate (laboratory) frame.
   ${\cal D}^{(J)}_{M \Omega }(\hat{r})$ is the rotational matrix element with
   $\hat{r}$ representing  the Euler angles for transformation from body-fixed to space-fixed frame.
   For $\Omega = 0$, ${\cal D}^{(J)}_{M \Omega }(\hat{r})$ reduces to spherical harmonic $Y_{JM}$.
     The dressed state of a bound level ($v, J$) coupled to a continuum can be written as
\bea \Psi_E &=& \sum_{M} \frac{\phi_{vJM}(r)}{r} \mid J \Omega M \rangle \mid e \rangle_{elec}  \nonumber \\
&+& \int d E' \beta_{E'} \sum_{\ell m_{\ell}} \frac{\psi_{ E' \ell m_{\ell}}(r)}{r} \mid \ell m_{\ell} 0 \rangle \mid g \rangle_{elec} \eea
where $ \phi_{vJM}(r)$ is the radial part of the bound state, $\mid \alpha \rangle_{elec}$
 denotes internal electronic part of the excited ($\alpha \equiv e$) and ground ($\alpha \equiv g$) molecular states,
 $\psi_{ E' \ell m_{\ell}}(r)$ represents energy-normalized partial wave scattering state with collision energy $E'$.
  Here $|\beta_{E'}|^2$ denotes density of states of the unperturbed continuum.  The internal electronic states $\mid \alpha \rangle$ are the functions of electronic coordinates and have parametrical dependence on the internuclear separation $r$. In electric dipole approximation, the interaction Hamiltonian is
$ H_{int} = \sum_{i=1,2} E_{L} \hat{\pi}.\hat{d}_i \label{hintatom}$
where $\hat{d}_i = - {\mathrm e} {\mathbf r}_i$ is the dipole moment of $i$-th atom
 whose  valence electron's position is given by ${\mathbf r}_i$ with respect to the center of mass of this atom.
  Here ${\mathrm e}$ represents an electron's charge, $E_L$ is the laser field amplitude and $\hat{\pi}$ is the polarization vector of the laser.  In the absence of hyperfine interactions, the total Hamiltonian in the center-of-mass frame of the two atoms can be written as
$ H = H_{elec}({\mathbf r_1},{\mathbf r_2};{\mathbf r_A},{\mathbf
r_B})  - \frac{\hbar^2}{2\mu}\nabla_r^2 - \frac{\hbar^2}{2
M}\nabla_R^2  + H_{int}$. where $H_{elec}$  includes terms which
depend on electronic coordinates. Here ${\mathbf r_A}$ and ${\mathbf
r_B}$ represent the position vectors of the nuclei of atoms A and B,
respectively,  $\nabla_r$ and $\nabla_R$ denote the Laplacian
operators corresponding to the relative coordinate ${\mathbf r} =
{\mathbf r_A} - {\mathbf r_B}$ and the center-of-mass coordinate
${\mathbf R} = ({\mathbf r_A} +{\mathbf r_B})/2$. From
time-independent Schr\"{o}dinger equation $H\Psi_{E} = E \Psi_{E}$,
using Born-Oppenheimer approximation we obtain the following coupled
equations
\bea
\left [ -\frac{\hbar^2}{2\mu}\frac{d^2}{d r^2} + B_J(r) + V_{e}(r) - \hbar \delta - E - i \hbar \frac{\gamma}{2} \right ]  \phi_{v J M} \nonumber \\
 = - \sum_{\ell m_{\ell}}\Lambda_{JM;\ell m_{\ell}} \tilde{\psi}_{E \ell m_{\ell}}
\label{coup1}\eea
\bea
\left [ - \frac{\hbar^2}{2\mu}\frac{d^2}{d r^2} + B_{\ell}(r) + V_{g}(r) - E \right]\tilde{\psi}_{E \ell m_{\ell}} \nonumber \\
= -\sum_{M}  \Lambda_{\ell m_{\ell}; J M} \phi_{v J M}
\label{coup2}
\eea
where  $B_J(r) = \hbar^2/(2\mu r^2) [J(J+1) - \Omega^2]$ is the
rotational term of excited molecular bound state and $B_\ell(r) =
\hbar^2/(2\mu r^2) \ell(\ell +1)$ is the centrifugal term in
collision of two ground state  atoms and $\tilde{\psi}_{E \ell
m_{\ell}}(r) =  \int_{E'} \beta_{E'} \psi_{E' \ell m_{\ell}}( r )d
E'$. The free-bound coupling matrix element is $\Lambda_{J M; \ell
m_{\ell}} =  \langle \ J M \Omega  \mid  \langle\phi_e( {\mathbf
r}_1, {\mathbf r}_2 ; r)\mid H_{int}\mid \phi_g( {\mathbf r}_1,
{\mathbf r}_2 ; r) \rangle \mid \ell m_{\ell} 0  \rangle$. The
molecular electronic wave functions $ \phi_{\alpha}( {\mathbf r}_1,
{\mathbf r}_2 ; r) = \langle {\mathbf r}_1,
{\mathbf r}_2 ; r  \mid \alpha \rangle$ can be
constructed from the symmetrized (or antisymmetrized) product of
atomic orbital  of the two atoms using Movre-Pichler model
\cite{MovreJPB13} which also provides the long-range part of
adiabatic potentials $V_{\alpha}$. We have here introduced a term
$\hbar \gamma/2$  corresponding to the natural line width of the
excited molecular state in order to take into account the inelastic
process of natural decay of the bound state.  $V_{e}(r)$  goes as $-
1/r^3$
  for $r \rightarrow \infty$ while  $ V_{g}(r)$ behaves as $- 1/r^6$ in the asymptotic regime. The excited state potential $V_e$ supports several bound states.
Here $ \delta = \omega_{L} - \omega_A $ is the frequency off-set
between  the laser frequency $\omega_L$ and atomic resonance
frequency $\omega_A$. The coupled Eqs. (\ref{coup1}) and
(\ref{coup2})  can be solved by the method of Green's function.  Let
$\phi_{vJ}^{0}$ be the bound state solution of the homogeneous part
($\Lambda = 0$) of (\ref{coup1}) with rovibrational energy $E_{vJ}$.
Note that we have here removed the subscript $M$ in the labelling of
wavefunctions for simplicity. The corresponding Green's function can
then be written as \be G_{v}(r,r')= -
\frac{\phi_{vJ}^{0}(r)\phi_{vJ}^{0}(r')}{\Delta E_{vJ} + i\hbar
\gamma/2 } \ee where $\Delta E_{vJ} = \hbar \delta + E - E_{vJ}$. We
can express the solution of equation (\ref{coup1}) in the form \be
\phi_{vJM} (r) = \int_{E'} d E'  \sum_{\ell m_{\ell}} A_{JM; \ell
m_{\ell}} \beta_{E'}\phi_{vJ}^{0}(r) \label{coupb}\ee where \bea
A_{JM; \ell m_{\ell}} = \frac{ \int dr'\Lambda_{JM; \ell
m_{\ell}}(r')\phi_{vJ}^{0}(r') \psi_{E \ell m_{\ell}}(r')}{\Delta
E_{vJ} + i\hbar \gamma/2 }\label{aeq} \label{VA}\eea The Green's
function for the homogeneous part of (\ref{coup2}) can be
constructed from the scattering soultions. Let $\psi_{E \ell}^{0 ,
reg}(r)$ and $\psi_{E \ell}^{0 , irr}(r)$ represent the regular and
irregular scattering solutions  in the absence of laser field.
$\psi_{E \ell}^{0 , reg}(r)$ vanishes at $r = 0 $ while  $\psi_{E
\ell}^{0 , irr}(r)$ is defined by boundary condition  at
$r\rightarrow\infty$ only. Asymptotically, they behave as $\psi_{E
\ell}^{0 , reg}(r) \sim  j_{\ell}\cos\eta_{\ell}
-n_{\ell}\sin\eta_{\ell}$ and $\psi_{E \ell}^{0 , irr}(r)  \sim
n_{\ell}\cos\eta_{\ell} +j_{\ell}\sin\eta_{\ell}$, where
$\eta_{\ell}$ is the background phase shift of $\ell$-th partial
wave in the absence of light field and $j_{\ell}$ and $ n_{\ell}$
are spherical Bessel and Neumann functions. According to threshold
laws, as $k \rightarrow 0$  we have $\eta_{\ell}  \sim k^{2 \ell +
1}$ for $\ell \le (n-3)/2$, otherwise  $\eta_{\ell} \sim
k^{n-\ell}$; with $n$ being the exponent of the inverse power-law
potential  at large separation. The Green's function \cite{mottmassey} for the
scattering wave function can be written as \bea
{\cal K}_{\ell} (r,r') &=& - \pi [\psi_{E \ell}^{0 ,reg}(r)\psi_{E \ell}^{0 ,irr}(r') + i \psi_{E \ell }^{0 , reg}(r)\psi_{E \ell }^{0 , reg}(r')] \nonumber \\
&&\hspace{5.0cm} r' > r \eea
 \bea
 {\cal K}_{\ell} (r,r') &=& - \pi [\psi_{E \ell}^{0 , reg}(r')\psi_{E \ell}^{0 , irr}(r)+i\psi_{E \ell}^{0  ,reg}(r)\psi_{E \ell}^{0 , reg}(r')] \nonumber  \\
 &&\hspace{5.0cm} r' < r   \eea

Substituting equation (\ref{coupb}) into equation (\ref{coup2})
and using ${\cal K}_{\ell} (r,r')$, we have \bea
\psi_{E \ell m_{\ell}}( r ) &=& \exp({i\eta_{\ell}})\psi_{E\ell}^{0} + \sum_{\ell' m_{\ell'} M } A_{ J M; \ell' m_{\ell'}}(E) \nonumber \\
&\times& \int {\cal K_{\ell}}(r,r')\Lambda_{ \ell m_{\ell}; J
M}(r')\phi_{vJ}^{0}(r')dr' \label{nscf}\eea where $\psi_{E\ell}^{0} = \psi_{E \ell}^{0  ,reg}$.
On substitution of
equation (\ref{nscf}) into (\ref{VA}) and after some algebra, we
obtain
\bea A_{J,M; \ell, m_{\ell}} &=&  \left [ f_{J, M; \ell m_{\ell}} + \left ( E_{J \ell}^{shift} - i \hbar \Gamma_{J\ell}/2 \right ) \tilde{A}_J \right ] \nonumber \\
&\times& \frac{1}{\Delta E_{vJ} + i \hbar \gamma/2}, \eea \bea
E_{J \ell}^{{\mathrm shift}} = &&\int\int d r' d r
\phi_{vJ}^{0}(r') \Lambda_{J M; \ell m_{\ell}}(r'){\rm Re}[{\cal K_{\ell}}(r',r)] \nonumber \\
&\times&   \Lambda_{\ell m_{\ell}; J M}(r)\phi_{vJ}^{0}(r).
\label{shift} \eea where $f_{J M; \ell m_{\ell}} =
\exp({i\eta_{\ell}}) \int\phi_{vJ}^{0}(r')
 \Lambda_{J M; \ell m_{\ell}}(r') \psi_{E\ell}^{0}(r') d r'$, $
\Gamma_{J\ell} = \frac{2 \pi }{\hbar} \left | \int
\phi_{vJ}^{0}(r)\Lambda_{J M; \ell m_{\ell}}(r)\psi_{E\ell}^{0}(r)d
r \right |^2$ and $\tilde{A}_J = \sum_{\ell, m_{\ell}, M} \frac{
f_{J M; \ell m_{\ell}}}{\Delta E_{vJ} + i \hbar \gamma/2 -
E_{J}^{shift} + i \hbar \Gamma_{J}/2 } $. Here $E_J^{shift} =
\sum_{\ell, m_{\ell}, M} E_{J \ell}^{shift}$ and $\Gamma_J =
\sum_{\ell, m_{\ell}, M} \Gamma_{J\ell}$. Using the asymptotic
boundary conditions of regular and irregular scattering wave
functions, the scattering $T$-matrix in the presence of light field
can now be written as \bea &&T_{\ell} = \frac{1}{2i} \left
[\exp({2i\eta_{\ell}})-1 \right ] - \exp[ 2 i\eta_{\ell}] \nonumber
\\ &\times&\frac{\sum_{\ell' M} f_{\ell m_{\ell}; J M} f_{J M; \ell'
m_{\ell'}}  }{\hbar\delta + E - E_{vJ} + i\hbar\gamma/2 -
E_{J}^{{\mathrm shift}} + i \hbar\Gamma_{J}/2} \label{tmat}\eea The
partial wave S-matrix element can now  be obtained from the relation
$S_{\ell} = 1 + 2 i T_{\ell}$. Only the second term on right hand
side (RHS) of the above equation contains the effect of light field.
This equation reveals that the T-matrix element for a partial wave
$\ell \ne 0$ can be modified by its indirect coupling with $\ell =
0$ via the excited rotational state. The modification for $\ell = 1$
is mainly due to the two-photon transition amplitude $t_{\ell' \ell}
= f_{\ell'=0 \rightarrow J=1} f_{J=1 \rightarrow \ell = 1}$. The
shift of Eq. (\ref{shift}) involves the real part of the Green's function ${\cal
K}_{\ell} (r, r')$ and the bound wave functions at two space
points $r$ and $r'$. Thus the shift depends on the radial correlation between
continuum and bound states. In the limit $k \rightarrow 0$, the
shift becomes independent of collision energy for all partial waves.
For large $k$ the shift will be vanishingly small.

For numerical illustration, we  consider a model system of two cold
ground state Na atoms coupled to $v = 48$ vibrational state of 1$_g$
molecular potential by a laser field. Recently,  several (up to $J =
6$) sharp rotational lines of this vibrational state have been
observed in PA spectra with a strong laser field \cite{GOMEZPRA75}.
The outer turning point of the excited $v = 48, J = 1$ level lies
inside the centrifugal barrier of $\ell \ne 0$ of the ground
continuum. Therefore,  nonzero partial waves are not expected to
contribute significantly to the PA transition amplitude at low
temperature in the weak-coupling regime. This situation can be
contrasted to the PA spectroscopy of higher rotational states where
transitions occur outside the barrier region \cite{bigelow}. The
results of our numerical calculations as tabulated  in
 Table-I  show that
the light shift $E_{J}^{{\mathrm shift}}$ can exceed the stimulated
line width $\Gamma_{J}$ by  more than one order of magnitude when
the PA laser intensity is as high as 10 kW/cm$^2$. The natural line
width $\gamma$ of the ro-vibrational states is of the order of 100
kHz \cite{spon}. However,  the light shift remains much smaller than
the rotational energy spacings $\Delta_J = E_{v J + 1} - E_{v J}$.
\begin{table}
\caption{ Numerically calculated partial energy shifts $E_{J
\ell}^{{\mathrm shift}}$ and partial stimulated line width
$\Gamma_{J \ell}$ (in unit of MHz) for PA laser intensity $I$ = 10
kW/cm$^2$ and collision energy $E = 50 \mu$K. Also given are the
rotational energy spacings $\Delta_J = E_{v J + 1} - E_{v J}$ (in
unit of GHz) for a few lowest $J$ values. The total shift and
stimulated line width for $J = 1$ are $E_1^{{\mathrm shift}} =
62.70$ MHz, \hspace{0.2cm} $\Gamma_1 = 2.87$ MHz, respectively.}
\begin{tabular}{c  c  c  c  c  c  c}
\hline
 \multicolumn{1}{c}{ $J$ } & \multicolumn{1}{c}{$\ell$} &
\multicolumn{1}{c}{$E_{J \ell}^{ {\mathrm shift }}$(MHz) }  &
\multicolumn{1}{c}{$\Gamma_{J \ell}$(MHz)} & \multicolumn{1}{c}{}
& \multicolumn{1}{c}{ $J$ }
& \multicolumn{1}{c} {$\Delta_J$(GHz)}   \\
\hline
1 &  0  &  -14.22  &  2.66 &  \vline &  1 &  1.56 \\
1 &  1  &  -17.20  &  0.21 &  \vline &  2 &  2.63 \\
1 &  2  &  -15.28 &   $10^{-4}$  &  \vline  & 3 &  3.78 \\
1 &  3  &  -16.00 &  $10^{-8}$  &  \vline  & 4 & 4.48  \\
 \hline
  \end{tabular}
\label{tb1}
\end{table}
Since the background (without laser) phase-shift $\eta_{\ell}
\rightarrow 0$ as the collision energy $E \rightarrow 0$, we can
approximate $t_{\ell,\ell'} \simeq \sqrt{\Gamma_{J \ell} \Gamma_{J
\ell'}}/2$ as real quantity unless the laser introduces a phase. It
then follows that the elastic scattering will be predominant if the
condition $
 (\Delta E_{vJ}  - E_{J}^{{\mathrm shift}}) >\!> \hbar (\gamma +
 \Gamma_{J})$ is fulfilled. As $k \rightarrow 0$, in the leading
 order in $k$ the ratio of the laser-induced change in p-wave
 $T$-matrix element  to that in s-wave one is given
 by $\sqrt{\Gamma_{J=1 \ell=1}/\Gamma_{J=1 \ell =0}} \simeq 0.28$.
  We can write an
energy-dependent $\ell$-wave scattering length as $a_{\ell} = -
\frac{T_{\ell}}{k} = a_{\ell}^{0} + a_{\ell}^{L}$, where
$a_{\ell}^{0}$ is the background scattering length and
$a_{\ell}^{L}$ denotes the laser-modified part of $a_{\ell}$. Note
that the scattering length $a_{\ell}$ ( for $\ell \ne 0$) as defined
here differs from the standard definition in scattering theory
\cite{newton}. However, $a_{\ell}$  as defined here can be related
to the standard $\ell$-wave scattering length  by using the behavior
of $a_{\ell}$ in the limit $k \rightarrow 0$ and thereby can be
compared with the results of Ref. \cite{pra69}. When $\delta_{vJ} =
\hbar \delta + E - E_{v J} \ge  0$, the real part of $a_{\ell}^{L}$
is positive since $E_{J}^{shift} < 0$. This implies that when PA
laser is tuned on resonance or on the blue side of the resonance,
the modified two-body interaction is repulsive. On the other hand,
when $\delta_{vJ} < E_{J}^{shift}$, the real part of $a_{\ell}^{L}$
is negative. This means that when PA laser is tuned on the red side
of the resonance by an amount exceeding $|E_{J}^{shift}|$, the
modified interaction becomes attractive. For the parameters given in
Table-I and assuming $\delta_{vJ} = - 15 \times \hbar \Gamma_1 $, we
make an estimate of $a_{\ell = 1}^{L}/a_{\ell = 1}^{0} \simeq 11$.
From low energy behavior of unperturbed scattering wave functions,
it follows that $\Gamma_{\ell} \sim k^{2 \ell + 1}$ for $\ell = 1$.
Therefore, $a_{\ell}^{L} \sim k^{\ell}$ as $k \rightarrow 0$, which
is significantly different from the behavior of background
scattering length $a_{\ell}^{0} \sim k^{2 \ell}$. This clearly
demonstrates that  nonzero partial wave scattering amplitudes can be
significantly modified by OFR. The modification of p-wave scattering
state can be experimentally observed by sending a weak probe laser
beam tuned near $J = 3$ transition while keeping the intense laser
beam (tuned near $J = 1$) operational as shown in Fig.1. Since $J =
3$ level can not be populated by PA transition from s-wave
scattering state, the appearance of $J = 3$ line in PA spectra will
unambiguously reveal optically induced p-wave Feshbach resonance.
The modification can also be enhanced by applying another intense
laser field tuned near $J = 3$ level as schematically shown with the
double blue arrow (on the right) in Fig. 1.

In conclusion, we have demonstrated that not only s-wave but also
higher partial wave atom-atom interaction can be manipulated by the
method of optical Feshbach resonance with an intense PA laser. We
have given quantitative estimate of relative modification of p-wave
scattering amplitude with a model calculation without hyperfine
interaction. However, inclusion of hyperfine interaction will not
alter the qualitative nature of our main results which are: (1) As a
result of strong-coupling PA laser-induced large light-shifts, atoms
experience dispersive light force leading to modified atom-atom
interaction. (2) PA laser-induced modification changes the threshold
behavior significantly.


\begin{references}

\bibitem{tiesingaPRA47} E. Tiesinga, B. J. Verhaar, and H. T. C.
Stoof, Phys. Rev. A {\bf 47}, 4114 (1993).

\bibitem{Ketterle} M. W. Zwierlein {\it et al.},
Nature {\bf 435}, 1047 (2005).

\bibitem{fermi} O'Hara {\it et al.},  Science {\bf 298}, 2179
(2002); M. Greiner,  C. A. Regal  and  D. S. Jin,  Nature {\bf
426}, 537 (2003); S. Jochim  {\it et al.}  Science {\bf 302}, 2101
(2003); M. W. Zwierlein  {\it et al.},  Phys. Rev. Lett. {\bf 91},
250401 (2003); M. Bartenstein  {\it et al.},  Phys. Rev. Lett.
{\bf 92}, 203201 (2004); J. Kinast {\it et al.},  Phys. Rev. Lett.
{\bf 92}, 150402 (2004);

\bibitem{bose} S. Inouye {\it et al.}, Nature {\bf 392}, 151 (1998);
 Ph. Courteille {\it et al.}, Phys. Rev. Lett. {\bf 81}, 69
(1998); J. L. Roberts {\it et al.}, Phys. Rev. Lett. {\bf 81},
5109 (1998).

\bibitem{fedichevPRL771} P. O. Fedichev, Y. Kagan, G. V. Shlyapnikov
 and J. T. M. Walraven, Phys. Rev. Lett. {\bf 77}, 2913 (1996).

\bibitem{fatemiPRL2002} F. K. Fatemi, K. M. Jones and P. D. Lett, Phys.
Rev. Lett. {\bf 85}, 4462 (2002).

\bibitem{theis} M. Theis {\it et al.} Phys. Rev. Lett. {\bf
93}, 123001 (2004);  G. Thalhammer {\it et al}, Phys. Rev. A {\bf
71}, 033403 (2005).

\bibitem{EnomotoPRL101} K. Enomoto, K. Kasa, M. Kitagawa
and Y. Takahashi, Phys. Rev. Lett. {\bf 101}, 203201 (2008).

\bibitem{parmp} For a recent review on PA, see K. M. Jones {\it et
al.}, Rev. Mod. Phys. {\bf 78}, 483 (2006).

\bibitem{ZhangPRA70} J. Zhang {\it et al},
Phys. Rev. A {\bf 70}, 030702 (2004).

\bibitem{rempePRA72} T. Volz {\it et al.}, Phys. Rev. A {\bf 72},
010704 (2005).

\bibitem{you} M. Marinescu and L. You, Phys. Rev. Lett. {\bf 81},
4596 (1998).

\bibitem{seidmanJCP103} T. Seidman, J. Chem. Phys. {\bf 103},
7887 (1995).

\bibitem{MovreJPB13} M. Movre and G. Pichler, J. Phys. B: Atom, Molec. Phys. {\bf 10}, 13 (1977).

\bibitem{mottmassey} N. F. Mott and H. S. W. Massey,  {\it Theory of Atomic Collisions}, 3rd ed. (Clarendon Press,
Oxford, 1965)

\bibitem{GOMEZPRA75} E. Gomez, A. T. Black, L. D. Turner, E. Tiesinga and P. D. Lett, Phys. Rev. A {\bf 75}, 013420 (2007).

\bibitem{bigelow} J. P. Shaffer, W. Chalupczak and N. P. Bigelow, Phys.
Rev. Lett {\bf 83}, 3621 (1999).

\bibitem{spon} K. M. Jones {\it et al.}, Phys. Rev. A {\bf 61},
012501 (1999).

\bibitem{newton} R. G. Newton, {\it Scattering Theory of Waves and Particles}, 2nd ed. (Springer, 1982).

\bibitem{pra69} C. Ticknor {\it et al.} Phys. Rev. A {\bf 69}, 042712 (2004).


\end{references}
\end{document}